\documentclass[twocolumn]{autart}    
\usepackage{harvard}

\usepackage{graphicx}          
\usepackage{amsmath,amssymb}
\usepackage{enumitem} 

\newcommand{\norm}[1]{\lVert #1 \rVert}

\newcommand{\abs}[1]{\lvert #1 \rvert}

\newcommand{\nnorm}[1]{{\vert\kern-0.25ex\vert\kern-0.25ex\vert #1 \vert\kern-0.25ex\vert\kern-0.25ex\vert}}

\usepackage[usenames]{xcolor}

\usepackage{mathtools}
\usepackage[colorlinks]{hyperref}
\usepackage{subcaption}

\def\deleted#1{\empty}
\def\changed#1#2{#2}

\usepackage{changepage}

\begin{document}

\begin{frontmatter}

\title{Stability of Markov regenerative switched linear systems\thanksref{footnoteinfo}} 

\thanks[footnoteinfo]{This paper was not presented at any IFAC 
meeting. Corresponding author M.~Ogura.}

\author[upenn]{Masaki Ogura}\ead{ogura@seas.upenn.edu}%
, 
\author[upenn]{Victor M.~Preciado}\ead{preciado@seas.upenn.edu}
\address[upenn]{Department of Electrical and Systems Engineering,
University of Pennsylvania, Philadelphia, PA 19104-6314, USA}


\begin{abstract}                          
In this paper, we give a necessary and sufficient condition for mean stability of switched linear systems having a Markov regenerative
process as its switching signal. This class of switched linear
systems, which we call Markov regenerative switched linear systems,
contains Markov jump linear
systems and semi-Markov jump linear systems as special cases. We show that a
Markov regenerative switched linear system is $m$th mean stable if and
only if a particular matrix is Schur stable, under the assumption that either $m$ is even or the system is positive.
\end{abstract}

\end{frontmatter}

\section{Introduction}

Among switched linear systems, those having a time-homogeneous Markov process as its switching signal, called Markov jump linear systems~\cite{Costa2013}, are of particular importance. One of the reasons for their importance is that time-homogeneous Markov processes are well suited to model stochastic phenomena presenting a constant rate of occurrence. Another reason is that the analysis and synthesis of Markov jump linear systems can be performed in a rather similar way to those for linear time-invariant systems by the introduction of auxiliary variables~\cite{Costa2013}.

However, it is often restricting to assume that the switching signals present constant transition rates, or even the Markovian property. To overcome this restriction, we find a wide variety of alternative switching signals in the literature \cite{Hou2005,Antunes2013,Ogura2014d,Ogura2013f}. A natural extension to time-homogeneous Markov processes are time-homogeneous semi-Markov processes, which are Markovian-like processes with time-varying transition rates \citeaffixed{Cinlar1975}{see}. The stability analysis of the corresponding switched linear systems can be found in~\citeasnoun{Antunes2013} and \citeasnoun{Ogura2013f}. Another extension are regenerative processes {\cite{Smith1955}}, which are, roughly speaking, stochastic processes that can be obtained by concatenating independent and identically distributed random functions (see \citeasnoun{Sigman1993} for the details). The mean stability analysis of linear systems subject to regenerative switchings is performed in~\citeasnoun{Ogura2014d}.

In this paper, we extend the works in \citeasnoun{Antunes2013}, \citeasnoun{Ogura2014d}, and \citeasnoun{Ogura2013f} to analyze the stability of \emph{Markov regenerative switched linear
systems}, which are switched linear systems whose switching signal is a
\emph{Markov regenerative process} (also called a \emph{semi-regenerative}
process) {\cite{Cinlar1975,Choi1994}}. Markov
regenerative processes form a large class of stochastic processes which
contains as special cases all the Markov, semi-Markov, and regenerative
processes. We show that {exponential} $m$th mean stability of a Markov
regenerative switched linear system is characterized by the spectral radius of a
matrix, under the assumption that either $m$ is even or the system is positive.
Extending various results in the literature
\cite{Antunes2013,Ogura2014d,Ogura2013f,Fang2002c}, the obtained
result enables us to analyze the stability of, for example, state-feedbacked
Markov jump linear systems with periodically observed mode signals (see a
discrete-time setting in~\citeasnoun{Cetinkaya2014b}), as well as controlled system
with failure-prone controllers with only one repairing
facility~\cite[Section~3.2]{Distefano2013}.

\ifdefined\supplemental
\pagestyle{fancy}
\fi

The paper is organized as follows. After introducing necessary notations, in
Section~\ref{sec:regsystems} we introduce the class of Markov regenerative
switched linear systems under consideration and then state a necessary and
sufficient condition for their exponential mean stability.
\deleted{Section~{\ref{sec:proof}} is devoted to the proof of our results.}
{Then, in Section~\ref{sec:application}, we present various applications of
the main result.} {The notation used in this paper is standard.} When $x\in
\mathbb{R}^n$ is nonnegative entrywise we write $x\geq 0$. The standard
Euclidean norm on $\mathbb{R}^n$ is denoted by $\norm{\cdot}$. Let $I$ and $O$
denote the identity and zero matrices, respectively. {The block diagonal
matrix with block diagonals $A_1$, $\dotsc$, $A_N$ is denoted by
$\bigoplus_{i=1}^N A_i$.} {The kronecker product of two matrices $A$ and $B$
is denoted by $A\otimes B$.} We say that $A \in \mathbb{R}^{n\times n}$ is Schur
stable if $A$ has the spectral radius less than one. Also we say that $A$ is
Hurwitz stable if all the eigenvalues of $A$ have negative real parts. For an
integrable random variable $X$, its expected value is denoted by $E[X]$ and its
conditional expectations by $E[X\mid \cdot]$.

\section{Stability characterization} \label{sec:regsystems}

This section introduces the class of Markov regenerative switched linear systems
and then presents a necessary and sufficient condition for their stability. We
need to first recall the definition of Markov renewal
processes~\cite{Cinlar1975}. {Let $N$ be a positive integer.} A stochastic
\changed{chain}{process} $(\theta, \tau) = \{(\theta_k, \tau_k)\}_{k\geq 0}$
taking values in $\{1, \dotsc, N\}\times [0, \infty)$ and satisfying $0 = \tau_0
\leq \tau_1 \leq \cdots$ is called a Markov renewal process~\cite{Cinlar1975} if
\begin{equation*}
\begin{multlined}
P(\theta_{k+1} = j, \tau_{k+1}-\tau_k \leq t \mid \theta_k, \dotsc, \theta_0, \tau_k, \dotsc, \tau_0)  
\\
=P(\theta_{k+1} = j, \tau_{k+1}-\tau_k \leq t \mid \theta_k)
\end{multlined}
\end{equation*}
holds for every $k$ and $t\geq 0$. We assume that $(\theta, \tau)$ is
time-homogeneous, that is, for all $i, j$ and $t \geq 0$, the probability
$P(\theta_{k+1} = j, \tau_{k+1}-\tau_k \leq t \mid \theta_k=i)$ is independent
of $k$. We note that, in this case, $\theta$ is a time-homogeneous Markov chain
and therefore has the constant transition probabilities $P(\theta_{k+1} = j \mid
\theta_k = i)$. In this paper, we furthermore assume that
\begin{equation}\label{eq:tauk+1-tauk>0}
\tau_{k+1}-\tau_k > 0
\end{equation}
with probability one for every $k$. Then, we can state the definition of Markov
regenerative processes as follows. 

\begin{defn}[{\citeasnoun{Cinlar1975}, \citeasnoun{Choi1994}}]
Let $\sigma = \{\sigma_t\}_{t\geq 0}$ be a stochastic process taking values in a
finite set~$\Lambda$. We say that $\sigma$ is \emph{Markov regenerative} if
there exists a Markov renewal process $(\theta, \tau)$ {taking values in
$\{1, \dotsc, N\} \times [0, \infty)$ and} satisfying the following conditions
for every $k$:
\begin{enumerate}[label={(P\arabic*) },ref={(P\arabic*)},leftmargin=*]
\item \label{item:stopping} $\tau_k$ is a stopping time for $\sigma$;

\item \label{item:determined } \deleted{$\theta_k$ is determined by
$\{\sigma_s\}_{s\leq \tau_k}$} {There exists a function $\pi \colon \Lambda
\to \{1, \dotsc, N\}$ such that $\pi(\sigma_{\tau_k}) = \theta_k$};

\item \label{item:Mar.regen} For {$j\in \{1, \dotsc, N\}$, }$\ell\geq 1$, $0\leq t_1 < t_2<
\cdots < t_\ell$, and a function $f \colon \Lambda^\ell \to [0,
\infty)$,
\begin{equation*}
\begin{multlined}
E[f(\sigma_{\tau_k + t_1}, \dotsc, \sigma_{\tau_k + t_\ell}) 
\mid 
\theta_k = j, \sigma_s; s\leq \tau_k] 
 \\
=E[f(\sigma_{t_1}, \dotsc, \sigma_{t_\ell}) \mid \theta_0 = j].
\end{multlined}
\end{equation*} 
\end{enumerate}
We call $\tau$ the \emph{regeneration times} of $\sigma$, and $(\theta, \tau)$
the \emph{embedded Markov renewal process} of $\sigma$.
\end{defn}

Among the three conditions in the definition, \ref{item:Mar.regen} is the most
important. It implies that, as far as prediction is concerned, at the
regeneration time $\tau_k$, the past information of the process
$\{\sigma_s\}_{s\leq \tau_k}$ is irrelevant {and only the value of
$\theta_k$ is needed}. Actually, \deleted{the definition} {\ref{item:Mar.regen}}
implies that
\begin{equation}\label{eq:probDistributionGivenTheta}
P(\sigma_{\tau_k + t} = \lambda \mid \theta_k = j, \sigma_s ; s\leq \tau_k) 
= 
P(\sigma_t = \lambda \mid \theta_0 = j)
\end{equation} 
for all $\lambda \in \Lambda$, $i, j\in\{1, \dotsc, N\}$, $t\geq 0$, and $k\geq 0$. {Equation \eqref{eq:probDistributionGivenTheta} indicates that, given $\theta_k = j$, the process~$\sigma$ ``regenerates'' at time $\tau_k$ as if it starts from time $0$, given $\theta_0 = j$.} {Moreover, $\sigma$ possesses a certain Markovian property at the regeneration time $\tau_k$; once we know the value of $\sigma_{\tau_k}$, using \ref{item:determined } we can determine the value $\theta_k$, which then determines the future distribution of $\sigma$ by \eqref{eq:probDistributionGivenTheta}.} {As is pointed out in~\citeasnoun{Cinlar1974}, it is convenient to take $\pi$ to be the identity, but this is not necessary.} Also we remark that \ref{item:stopping} is a technical condition needed to state \deleted{{\ref{item:determined }} and }\ref{item:Mar.regen}. For the details, the readers are referred to \citeasnoun{Cinlar1975} and \citeasnoun{Cinlar1974}.

Then, we introduce the class of switched linear systems studied in this paper.

\begin{defn}
Let $\sigma$ be a Markov regenerative process and let $A_\lambda \in
\mathbb{R}^{n \times n}$ for each $\lambda \in \Lambda$. Then, the stochastic
differential equation
\begin{equation}\label{eq:def:Sigma}
\Sigma :
\frac{dx}{dt} = A_{\sigma_t} x, 
\end{equation}
where $x(0) = x_0 \in \mathbb{R}^n$ is a constant, is called a
\emph{Markov regenerative switched linear system}.
\end{defn}

\deleted{Markov jump linear systems {\cite{Costa2013}}, semi-Markov jump linear
systems [1, 13], and regenerative switched linear systems {\cite{Ogura2014d}}
are all Markov regenerative switched linear systems because any of
time-homogeneous Markov, time-homogeneous semi-Markov, and regenerative
processes are Markov regenerative {\cite{Cinlar1975}}.} For example, a
time-homogeneous Markov process $r$ is Markov regenerative with an underlying
embedded Markov renewal process being $\{(r_{t_k}, t_k)\}_{k\geq 0}$, where $0 =
t_0 < t_1 < t_2 < \cdots$ are the times at which the process $r$ changes its
value. {Therefore, Markov jump linear systems~\cite{Costa2013} are Markov
regenerative switched linear systems.} \deleted{Also, a regenerative process is by
definition a semi-regenerative process with $\Lambda$ being a singleton.}
{In Section~\ref{sec:application}, we will show in detail that more general
classes of switched linear systems, such as semi-Markov jump linear
systems~\cite{Antunes2013,Ogura2013f} and regenerative switched linear
systems~\cite{Ogura2014d}, are contained in the class of Markov regenerative
switched linear systems.}

Based on the embedded Markov renewal process, we can naturally define
the stability of Markov regenerative switched linear systems as
follows.

\begin{defn}\label{defn:stbl}
Given a positive integer~$m$, we say that $\Sigma$ is
\emph{exponentially $m$th mean stable} if there exist $C>0$ and $\beta
> 0$ such that $E[\norm{x(t)}^m] \leq Ce^{-\beta t}\norm{x_0}^m$ for
every $x_0$ and $\theta_0$.
\end{defn}

Also, we here introduce the positivity of $\Sigma$. 

\begin{defn}
We say that $\Sigma$ is \emph{positive} if $x_0 \geq 0$ implies $x(t) \geq 0$
with probability one for \changed{every}{all $\theta_0$ and} $t \geq 0$.
\end{defn}

We remark that, for $\Sigma$ to be positive, it is clearly sufficient that all the matrices $A_\lambda$ ($\lambda \in \Lambda$) are Metzler, i.e., the off-diagonal entries of each $A_\lambda$ are all nonnegative~\cite{Farina2000}. However, this sufficient condition is not necessary {as is shown in} \citeasnoun[Example~10]{Ogura2014d} for regenerative switched linear systems.

In order to state the main result of this paper, we recall the notion of induced
matrices. First we define \citeaffixed{Parrilo2008}{see, e.g.,} the $m$-lift of~$x \in
\mathbb{R}^n$, denoted by~$x^{[m]}$, as the real vector of length~$n_m =
\binom{n+m-1}{m}$ with its elements being the lexicographically ordered
monomials~$\sqrt{\alpha!}\,x^\alpha$ ($\alpha! := {m!}/(\alpha_1! \dotsm
\alpha_n!)$) that are indexed by all the possible exponents $\alpha = (\alpha_1,
\dotsc, \alpha_n)$ summing up to $m$. {For example, for $x = [x_1, x_2]^\top$ we have $x^{[1]} = x$, $x^{[2]} = [x_1^2, \sqrt 2 x_1 x_2,
x_2^2]^\top$, and $x^{[3]} = [x_1^3, \sqrt 3 x_1^2 x_2, \sqrt 3 x_1 x_2^2,
x_2^3]^\top$.} \label{eq:ex:lifts} Then, we define the $m$th induced matrix of
$A$, denoted by $A^{[m]}$, as the $n_m \times n_m$ unique
matrix~\cite{Parrilo2008} satisfying $(Ax)^{[m]} = A^{[m]} x^{[m]}$ for every
$x\in\mathbb{R}^n$.

The next theorem is the main result of this paper.

\begin{thm}\label{theorem:main}
Let $p_{ij} = P(\theta_1 = j \mid \theta_0 = i)$ for all $i, j\in \{1,
\dotsc, N\}$. For all $0\leq s\leq t < \infty$, define the
$\mathbb{R}^{n\times n}$-valued random variable $\Phi(t;s)$ by the
differential equation ${\partial \Phi}/{\partial t} = A_{\sigma_t}
\Phi(t;s)$ with the initial conditions $\Phi(t; t) = I_n$ for every
$t\geq 0$. Assume that the following two conditions hold:
\begin{enumerate}[label={(A\arabic*) },ref={(A\arabic*)}, leftmargin=*]
\item \label{item:evenORnon-neg} Either $m$ is even or $\Sigma$ is
positive; 

\item \label{item:hk<T} There exists $T>0$ such that $\tau_{k+1} -
\tau_k \leq T$ with probability one for every $k\geq 0$.
\end{enumerate}
Then, $\Sigma$ is exponentially $m$th mean stable if and only if the $(Nn_m)
\times (Nn_m)$ real  block matrix $\mathcal A = [\mathcal A_{ij}]_{1\leq i,
j\leq N}$ with the $(i,j)$-block $\mathcal A_{ij}$ being defined by
\begin{equation}\label{eq:calAij}
\mathcal A_{ij} = p_{ji} E[\Phi(\tau_1; 0)^{[m]} \mid \theta_0 = j, \theta_1 = i]
\end{equation}
is Schur stable.
\end{thm}

{In Section~\ref{sec:application}, we will show that the above theorem can recover stability characterizations of Markov jump linear systems~\cite{Fang2002c,Ogura2013f}, semi-Markov jump linear systems~\cite{Antunes2013,Ogura2013f}, and regenerative switched linear systems~\cite{Ogura2014d}. We will also show that the theorem can give a stability condition for other switched linear systems whose stability cannot be studied using the methods in the literature.}

\subsection{Proof} \label{sec:proof}

We present a proof of Theorem~\ref{theorem:main} in this subsection. This proof consists of the following two steps. We first show, in Proposition~\ref{prop:SSigma}, that the stability of $\Sigma$ can be analyzed based on its discretized version. The proof of the proposition utilizes the technique developed in~\citeasnoun{Ogura2014d} to study regenerative switched linear systems. To analyze the discretized version of the system, we then present Proposition~\ref{prop:stab:mu:char}, which extends Theorem~3.4 in \citeasnoun{Ogura2013f} for not necessarily positive systems. \label{difference}

Let $\Sigma$ be a Markov regenerative switched linear system. Then, the
discretized process $x_d = \{x(\tau_k)\}_{k\geq 0}$ is the solution of the
discrete-time system $\mathcal S \Sigma: x_d(k+1) =
\Phi(\tau_{k+1};\tau_k)x_d(k)$. In order to proceed, we introduce a class of
switched linear systems called discrete-time semi-Markov jump linear
systems~\cite{Ogura2013f}. Let $\{F_k\}_{k\geq 0}$ be a stochastic process
taking values in $\mathbb{R}^{n\times n}$. The system $\Sigma_d: x_d(k+1) =
F_kx_d(k)$ ($x_d(0) = x_0$) is said to be a \emph{discrete-time semi-Markov jump
linear system} if there exists a time-homogeneous Markov chain~$\theta$ taking
values in $\{1, \dotsc, N\}$ such that, for all $k\geq 0$, $i, j \in \{1,
\dotsc, N\}$, and a Borel subset~$G$ of $\mathbb{R}^{n\times n}$, there holds
that
\begin{equation}\label{eq:semiMarkov}
\begin{multlined}
P(\theta_{k+1} = j, F_k \in G \mid \theta_k, \dotsc, \theta_0,
F_{k-1}, \dotsc, F_0) \\= P(\theta_{k+1} = j, F_k \in G \mid \theta_k), 
\end{multlined}
\end{equation}
and the conditional probability 
\begin{equation}\label{eq:condidion}
P(\theta_{k+1} = j,  F_k \in G \mid \theta_k =i)
\end{equation}
does not depend on $k$. The mean stability and the positivity of
$\Sigma_d$ is defined in the following standard manner. For a positive
integer $m$, we say that $\Sigma_d$ is \emph{exponentially $m$th mean
stable} if there exist $C>0$ and $\beta > 0$ such that
$E[\norm{x_d(k)}^m] \leq Ce^{-\beta k} \norm{x_0}^m$ for all $x_0$ and
$\theta_0$. Also we say that $\Sigma_d$ is \emph{stochastically $m$th
mean stable} if $\sum_{k=0}^\infty E[\norm{x_d(k)}^m]$ is finite for
all $x_0$ and $\theta_0$. Finally, $\Sigma_d$ is said to be
\emph{positive} if $x_0\geq 0$ implies $x_d(k) \geq 0$ with
probability one for every $k$ and $\theta_0$.

The next proposition relates the stability of $\Sigma$ and $\mathcal
S\Sigma$.

\begin{prop}\label{prop:SSigma}
$\mathcal S \Sigma$  is a discrete-time semi-Markov jump linear system. Moreover, if
\ref{item:hk<T} holds, then the following statements are true:
\begin{itemize}
\item If $\Sigma$ is exponentially {$m$th} mean stable, then $\mathcal S \Sigma$ is stochastically $m$th mean stable. 

\item If $\mathcal S \Sigma$ is exponentially {$m$th} mean stable, then so is $\Sigma$. 
\end{itemize}
\end{prop}

\begin{pf}
Let $F_k = \Phi(\tau_{k+1};\tau_k)$. Also, we let $(\Omega, \mathcal M, P)$
denote the underlying probability space. We denote the $\sigma$-algebra of
$\Omega$ generated by a set $X$ of random variables by $\mathcal M(X)$. We
define the following $\sigma$-algebras; $\mathcal M_1 = \mathcal
M(\{\theta_k\})$, $\mathcal M_2 = \mathcal M(\{\theta_k, \dotsc, \theta_0,
F_{k-1}, \dotsc, F_0\})$, $\mathcal M_3 = \mathcal M(\{\theta_k, \dotsc,
\theta_0, \sigma_s ;{s\leq\tau_k}\})$, and $\mathcal M_4 = \mathcal
M(\{\theta_k, \sigma_s; s\leq\tau_k\})$. Then, we can show
\begin{equation}\label{eq:1234}
\mathcal M_1 \subset \mathcal M_2 \subset \mathcal M_3 = \mathcal M_4.
\end{equation}
The first inclusion is obvious. The second inclusion is true because $F_0,
\dotsc, F_{k-1}$ are measurable on $\mathcal M_3$. Finally, the last identity
follows from \ref{item:determined }. Now, from {{\ref{item:Mar.regen}}}
we know that $P(\theta_{k+1} = j, F_k\in G \mid \mathcal M_1) = P(\theta_{k+1}
= j, F_k\in G \mid \mathcal M_4)$. Therefore, by \eqref{eq:1234} and
\citeasnoun[Lemma~1.1]{Ogura2013f}, we conclude that $P(\theta_{k+1} = j, F_k\in G
\mid \mathcal M_1) = P(\theta_{k+1} = j, F_k\in G \mid \mathcal M_2)$, which is
equivalent to \eqref{eq:semiMarkov}. Also, the conditional
probability~\eqref{eq:condidion} is independent of $k$ by \ref{item:Mar.regen}
and the time-homogeneity of~$\theta$. Therefore, $\mathcal S \Sigma$ is a
discrete-time semi-Markov jump linear system.

The proof of the second statement can be done in the same way as the proof for the implication $[2\Rightarrow 3]$ of \citeasnoun[Theorem~12]{Ogura2014d} due to \ref{item:hk<T} and the assumption \changed{that $\tau_{k+1}-\tau_k > 0$ with probability one}{\eqref{eq:tauk+1-tauk>0}}. Also, we can prove the third statement in the same way as the proof for $[3 \Rightarrow 1]$ of \citeasnoun[Theorem~12]{Ogura2014d} by \ref{item:hk<T}. The details of the proofs are thus omitted.
\end{pf}

Proposition~\ref{prop:SSigma} shows that the stability analysis of~$\Sigma$ could be reduced to the stability analysis of {discrete-time} semi-Markov jump linear systems. The next proposition gives a characterization of the {exponential $m$th} mean stability of {discrete-time} semi-Markov jump linear systems, extending \citeasnoun[Theorem~3.4]{Ogura2013f} to the case where $m$ is even.

\begin{prop}\label{prop:stab:mu:char}
For $i, j\in\{1, \dotsc, N\}$, let $p_{ij}$ denote the transition
probability of $\theta$ from $i$ to $j$. Assume that either $m$ is
even or $\Sigma_d$ is positive. Then, the following statements are
equivalent.
\begin{enumerate}
\item \label{item:mu:expsta} $\Sigma_d$ is exponentially $m$th
mean stable;

\item \label{item:mu:stosta} $\Sigma_d$ is stochastically
$m$th mean stable;

\item \label{item:mu:Shcsta} The $(Nn_m) \times (Nn_m)$ block matrix
$\mathcal F$ with the $(i,j)$-block  $\mathcal F_{ij}  = p_{ji}
E[F_0^{[m]} \mid \theta_0 = j, \theta_1 = i] \in \mathbb{R}^{n_m
\times n_m}$ is Schur stable.
\end{enumerate}
\end{prop}

For the proof of this proposition, we will need the next lemma. 

\begin{lem}\label{lemma:cone:R}
Assume that $m$ is an even integer. Let $K$ be the closed convex hull of
$(\mathbb{R}^n)^{[m]} = \{x^{[m]} : x\in\mathbb{R}^n\}$ in
$\mathbb{R}^{n_m}$. Then, there exists a norm~$\nnorm{\cdot}$ on
$\mathbb{R}^{Nn_m}$ and $f\in \mathbb{R}^{Nn_m}$ such that $\nnorm{x}
= f^\top x$ for every $x\in K\times \cdots \times K$.
\end{lem}

\begin{pf}
We first show that $K$ is a proper cone~\cite[Chapter~26]{Tam2006},
that is, $K$ is a closed and convex cone having a nonempty interior
and satisfying
\begin{equation}\label{eq:ptt}
K\cap (-K) = \{0\}, 
\end{equation}
where $-K := \{-x: x\in K\}$. $K$ is clearly a closed and convex cone.

Let us show \eqref{eq:ptt}. For each $i \in \{1, \dotsc, n\}$, let
$\ell_i$ denote the position of the monomial~$x_i^m$ in the
vector~$x^{[m]}$, i.e., we assume that $(x^{[m]})_{\ell_i} = x_i^m$.
Let us show the existence of a constant~$C>0$ such that every $y \in
K$ satisfies
\begin{equation}\label{eq:y_ell}
y_{\ell_i}\geq 0, \ 
\abs{y_\ell} \leq C\sum_{i=1}^n y_{\ell_i},
\end{equation}
for all $1\leq i\leq n$ and $1\leq \ell \leq n_m$. Take an arbitrary $y$ in the convex hull of $(\mathbb{R}^n)^{[m]}$. Then there exist $x_1, \dotsc, x_N \in \mathbb{R}^n$ and positive numbers $c_1, \dotsc, c_N$ such that $y = \sum_{j=1}^N c_j x_j^{[m]}$. Without loss of generality we can assume $c_j = 1$. Then, $y_{\ell_i} = \sum_{j=1}^N (x_j)_i^m \geq 0$ because $m$ is even. Next, let $1 \leq \ell \leq n_m$ be arbitrary and let $\alpha = (\alpha_1, \dotsc, \alpha_n)$ be the exponent of the monomial~$(x^{[m]})_\ell$, i.e., we suppose that $(x^{[m]})_\ell = \sqrt{\alpha!} x^\alpha$. Then, the inequality of arithmetic and geometric means implies $\abs{(x^{[m]})_\ell} \leq m^{-1}\sqrt{\alpha!} \sum_{i=1}^{n} \alpha_i x_i^m$. Hence, the triangle inequality shows that $\abs{y_\ell} \leq \sum_{j=1}^N \abs{(x_j^{[m]})_\ell} \leq m^{-1}\sqrt{\alpha!} \sum_{i=1}^n \sum_{j=1}^N (x_j)_i^m \leq C \sum_{i=1}^n y_{\ell_i}$, where $C  = m^{-1}\max_{\alpha}(\sqrt{\alpha!})$. Therefore, every $y$ in the convex hull of $(\mathbb{R}^n)^{[m]}$ satisfies \eqref{eq:y_ell}. A limiting argument thus shows that \eqref{eq:y_ell} is satisfied by every $y \in K$.  Now assume $y, -y \in K$. This implies $y_{\ell_i} = 0$ for every $i$ and therefore $y_\ell = 0$ for every $\ell$ by \eqref{eq:y_ell}. Hence~\eqref{eq:ptt} holds.

Finally, by \citeasnoun[Lemma~1.5]{Ogura2013f}, the difference $K-K = \{x-y:
x, y\in K\}$ equals the whole space~$\mathbb{R}^{n_m}$. This in fact
shows that the interior of $K$ is nonempty because, in general, a
closed and convex cone $K$ satisfying \eqref{eq:ptt} has a nonempty
interior if and only if the difference $K - K$ coincides with the
whole space~\cite[Chapter~26]{Tam2006}.

Now, since $K$ is a proper cone, the $N$-direct product $K\times
\cdots \times K$ in $\mathbb{R}^{Nn_m}$ is also a proper cone.
Therefore there exists \cite[Section~2]{Seidman2005} a norm
$\nnorm{\cdot}$ on $\mathbb{R}^{Nn_m}$ and a vector $f\in
\mathbb{R}^{Nn_m}$ having the desired property. This completes the
proof.
\end{pf}

We then prove Proposition~\ref{prop:stab:mu:char}:

\def\Elproofname{PROOF of Proposition~\ref{prop:stab:mu:char}.}
\begin{pf}
The case where $\Sigma_d$ is positive is shown in
\citeasnoun[Theorem~3.4]{Ogura2013f}. Let us assume that $m$ is even. We shall show
the cycle [\ref{item:mu:expsta} $\Rightarrow$ \ref{item:mu:stosta} $\Rightarrow$
\ref{item:mu:Shcsta} $\Rightarrow$ \ref{item:mu:expsta}]. It is easy to prove
[\ref{item:mu:expsta} $\Rightarrow$ \ref{item:mu:stosta}]. The implication
[\ref{item:mu:stosta} $\Rightarrow$ \ref{item:mu:Shcsta}] can be proved in the
same way as in the proof of Theorem 3.4 in \citeasnoun{Ogura2013f} without the
positivity assumption.

Let us prove [\ref{item:mu:Shcsta} $\Rightarrow$ \ref{item:mu:expsta}] assuming
$m$ is even. Take a norm $\nnorm{\cdot}$  on $\mathbb{R}^{Nn_m}$ and $f \in
\mathbb{R}^{Nn_m}$ satisfying the linearity property described in
Lemma~\ref{lemma:cone:R}. By the equivalence of the norms on a
finite-dimensional linear space, we can take a constant $C_1>0$ such that
$C_1^{-1}\norm{\cdot} \leq \nnorm{\cdot} \leq C_1\norm{\cdot}$. \deleted{Also define
the $\{0, 1\}$-valued discrete-time stochastic chains $\zeta_1, \dotsc, \zeta_N$
by $\zeta_i(k) = 1$ if and only if $\theta_k = i$ for all $i \in \{1, \dotsc,
N\}$ and $k\geq 0$.} {Let us consider the stochastic process $e_{\theta_k}$,
where $e_1$, $\dotsc$, $e_N$ denote the standard unit vectors in
$\mathbb{R}^N$.} Using \deleted{the definition of the processes $\zeta_i$ and} the
general identity $\norm{x^{[m]}} = \norm{x}^m$ \citeaffixed{Parrilo2008}{see}, we can
show the inequality $\norm{x_d(k)}^m = \norm{x_d(k)^{[m]}} = \norm{e_{\theta_k}
\otimes x_d(k)^{[m]}} \leq C_1\nnorm{e_{\theta_k} \otimes x_d(k)^{[m]}}$. Since
$e_{\theta_k} \otimes x_d(k)^{[m]} \in K\times \cdots \times K$, the linearity
of the norm $\nnorm{\cdot}$  shows that
\begin{equation}\label{eq:kronewqpre}
\begin{aligned}
E[\norm{x_d(k)}^m] 
&\leq
C_1\nnorm{E[e_{\theta_k} \otimes x_d(k)^{[m]}]}
\\
&\leq
C_1^2\norm{E[e_{\theta_k} \otimes x_d(k)^{[m]}]}. 
\end{aligned}
\end{equation}
On the other hand, by the identity
\begin{equation}\label{eq:one-step}
E[e_{\theta_{k+1}} \otimes x_d(k+1)^{[m]}] = \mathcal F 
E[e_{\theta_k} \otimes x_d(k)^{[m]}]
\end{equation} 
proved in
\citeasnoun[Proposition~3.8]{Ogura2013f}, if $\rho(\mathcal F) < 1$ then
there exist $C_2>0$ and $\beta>0$ such that
\begin{equation}\label{eq:kroneq}
\begin{aligned}
\norm{E[e_{\theta_k} \otimes x_d(k)^{[m]}] }
&\leq
C_2e^{-\beta k}\norm{e_{\theta_0} \otimes x_0^{[m]}}
\\
&=
C_2e^{-\beta k}\norm{x_0^{[m]}}
\\
&=
C_2e^{-\beta k}\norm{x_0}^m. 
\end{aligned}
\end{equation}
Therefore, the inequalities~\eqref{eq:kronewqpre} and \eqref{eq:kroneq}
prove that $\Sigma_d$ is exponentially $m$th mean stable. This
completes the proof of Proposition~\ref{prop:stab:mu:char}.
\end{pf}
\def\Elproofname{PROOF.}

Now we can readily prove Theorem~\ref{theorem:main}.
\ref{item:evenORnon-neg} implies that either $m$ is even or  $\mathcal
S \Sigma$ is  positive. Therefore, by
Proposition~\ref{prop:stab:mu:char} and the first statement of
Proposition~\ref{prop:SSigma}, we see that the following three
properties are equivalent; the exponential $m$th mean stability of
$\mathcal S\Sigma$, the stochastic $m$th mean stability of $\mathcal
S\Sigma$, and the Schur stability of $\mathcal A$. This equivalence
and also the second and third statements of
Proposition~\ref{prop:SSigma} immediately prove the main result of
Theorem~\ref{theorem:main}.

{
\section{Applications}\label{sec:application}
}

{In this section, we illustrate how Theorem~\ref{theorem:main}, the main
result of this paper, can be used to recover various stability characterizations
of switched linear systems derived in the literature
(Subsections~\ref{sec:regen}--\ref{sec:mjls}), as well as to analyze the
stability of systems that cannot be analyzed by current techniques in the literature
(Subsection~\ref{subsec:periodic}). Throughout this section, we will use the
following notation \citeaffixed{Brockett1973}{see}. For a
matrix $A\in\mathbb{R}^{n\times n}$, we define $A_{[m]} \in
\mathbb{R}^{n_m\times n_m}$ as the unique real matrix such that
\begin{equation}\label{eq:def:_[m]}
(\exp(At))^{[m]} = \exp(A_{[m]}t).
\end{equation}}

{\subsection{Regenerative switched linear systems}\label{sec:regen}}

{In this subsection, we discuss some implications of the stability characterization provided for regenerative switched linear systems~\cite{Ogura2014d}. We first recall the definition of regenerative processes. We say that a stochastic process $\sigma$ is a regenerative process~\cite{Smith1955} if there exists a random variable $R_1>0$ such that (\emph{i}) $\{\sigma_{t+R_1}\}_{t\geq 0}$ is independent of $\{\{\sigma_t\}_{t<R_1}, R_1\}$, and (\emph{ii}) $\{\sigma_{t+R_1}\}_{t\geq 0}$ has the same joint distribution as $\{\sigma_t \}_{t\geq 0}$. Then, following \citeasnoun{Ogura2014d}, we say that $\Sigma$ given in \eqref{eq:def:Sigma} is a {regenerative switched linear system} if $\sigma$ is a regenerative process.}

Let random variables $\{R_k \}_{k\geq 1}$ be independent and identically distributed and define $\tau_k = R_1 + \cdots + R_k$ for each $k\geq 1$. Then, $\sigma$ can be broken into independent and identically distributed cycles~$\{\sigma_t \}_{0\leq t < \tau_1}$, $\{\sigma_t \}_{\tau_1\leq t < \tau_2}$, $\dotsc$ \citeaffixed{Sigman1993}{see}. From this observation, it is immediate to see~\cite{Cinlar1975} that $\sigma$ is Markov regenerative process with its embedded Markov renewal process being $(\theta, \tau)$, where $\theta$ is the constant sequence~$\{1, 1, \dotsc\}$. Therefore, we can define the exponential $m$th mean stability of a regenerative switched linear system by Definition~\ref{defn:stbl}. We can furthermore prove, as a corollary of Theorem~\ref{theorem:main}, the following stability characterization for regenerative switched linear systems originally given in \citeasnoun[Theorem~12]{Ogura2014d}:

{\begin{cor}[{\citeasnoun[Theorem~12]{Ogura2014d}}]\label{cor:regen}
Let $\Sigma$ be a regenerative switched linear system. Assume that
\ref{item:evenORnon-neg} is true and, moreover, there exists $T>0$ such that
$R_1 \leq T$. Then, $\Sigma$ is exponentially $m$th mean stable if and only if
the matrix $E[\Phi(R_1; 0)^{[m]}]$ is Schur stable.
\end{cor}}

\begin{pf}
{We first remark that, since $R_1 > 0$, assumption~\eqref{eq:tauk+1-tauk>0}
is satisfied. Moreover, $R_1 \leq T$ guarantees that \ref{item:hk<T} is
satisfied. Since \ref{item:evenORnon-neg} is true by the assumption in the
corollary, we can apply Theorem~\ref{theorem:main}. Since $\theta \equiv 1$, the
matrix $\mathcal A$ given by \eqref{eq:calAij} equals the $n_m \times n_m$
matrix $p_{11} E[\Phi(\tau_1; 0)^{[m]} \mid \theta_0 = 1, \theta_1 = 1] =
E[\Phi(R_1; 0)^{[m]} ]$. Therefore, Theorem~\ref{theorem:main} readily
completes the proof of the corollary. }
\end{pf}%

\subsection{{Semi-Markov jump linear systems}}

{We now consider another class of switched linear systems called
semi-Markov jump linear systems~\cite{Ogura2013f}. Assume that $(\theta, \tau)$
is a Markov renewal process taking values in $\{1, \dotsc, N\} \times [0, \infty)$, whose
definition was given in the beginning of Section~\ref{sec:regsystems}. Then, the
stochastic process $\{\sigma_t \}_{t\geq 0}$ defined by
\begin{equation}\label{eq:semi-Markov}
\sigma_t = \theta_k,\ \tau_k\leq t< \tau_{k+1}, 
\end{equation}
is called a semi-Markov process~\cite{Cinlar1975}. Then, following \citeasnoun{Ogura2013f}, we say that $\Sigma$ given by \eqref{eq:def:Sigma} is a {semi-Markov jump linear system} if $\sigma$ is a semi-Markov process. It is easy to see that $\sigma$ is a Markov regenerative process with its embedded renewal process being $(\theta, \tau)$. The mapping $\pi$ in \ref{item:determined } is taken to be the identity. Without loss of generality, we can assume \eqref{eq:tauk+1-tauk>0}.}

The following corollary of Theorem~\ref{theorem:main}, which extends
\citeasnoun[Theorem~2.5]{Ogura2013f} to not necessarily positive systems, immediately
follows from \eqref{eq:def:_[m]} and \eqref{eq:semi-Markov}:

\begin{cor}\label{cor:semi-}
Suppose that $\Sigma$ is a semi-Markov jump linear system and assume that
conditions \ref{item:evenORnon-neg} and \ref{item:hk<T} hold. Then, $\Sigma$
is exponentially $m$th mean stable if and only if the $(Nn_m) \times
(Nn_m)$ real block matrix $\mathcal A = [\mathcal A_{ij}]_{1\leq i, j\leq
N}$ with the $(i,j)$-block $\mathcal A_{ij}$ being defined by $\mathcal
A_{ij} = p_{ji} E[\exp((A_j)_{[m]} \tau_1) \mid \theta_0 = j, \theta_1 = i]$
is Schur stable.
\end{cor}

\subsection{Markov jump linear systems}\label{sec:mjls}

In this subsection, we show that Theorem~\ref{theorem:main} can recover
stability characterizations~\cite{Ogura2013f,Fang2002c} for a class of switched
linear systems called Markov jump linear systems~\cite{Costa2013}. We say that
$\Sigma$ given by \eqref{eq:def:Sigma} is a {Markov jump linear system} if
$\sigma$ is a time-homogeneous Markov process. Generalizing the stability
characterizations of Markov jump linear
systems in \citeasnoun[Theorem~3.3]{Fang2002c} for $m=2$ and in
\citeasnoun[Theorem~5.1]{Ogura2013f} for positive systems, we can show the next corollary of Theorem~\ref{theorem:main}:

\begin{cor}\label{cor:Markov}
Suppose that $\Sigma$ is a Markov jump linear system and $\sigma$ can take
values in $\{1, \dotsc, N\}$. Assume that either $m$ is even or $A_1$, $\dotsc$,
$A_N$ are Metzler. Let $Q$ be the infinitesimal generator of Markov
process~$\sigma$. Then, $\Sigma$ is exponentially $m$th mean stable if and only
if the matrix $\mathcal B_\Sigma = Q^\top\otimes I_{n_m} + \bigoplus_{i=1}^N
(A_i)_{[m]}$ is Hurwitz stable. 
\end{cor}

\begin{pf} 
{Notice that, for $h>0$ arbitrarily chosen, $\sigma$ is a Markov
regenerative process with the embedded Markov renewal process $(\theta, \tau) =
\{(\sigma_{kh}, kh)\}_{k\geq 0}$ by the Markovian property of $\sigma$.
Therefore, $\Sigma$ is a Markov regenerative switched linear system. The
assumption~\eqref{eq:tauk+1-tauk>0} is trivially true and, also, condition
\ref{item:hk<T} is satisfied with $T = h$. Moreover, the assumption of the
corollary ensures that condition \ref{item:evenORnon-neg} also holds true. Let
$p_{ij}(h) = P(\theta_1 = j \mid \theta_0 = i) = P(\sigma_h = j \mid \sigma_0 =
i)$. Then, by Theorem~\ref{theorem:main}, $\Sigma$ is exponentially $m$th mean
stable if and only if the block matrix $\mathcal A(h) = [\mathcal
A(h)_{ij}]_{1\leq i, j\leq N}$ given by $\mathcal A(h)_{ij} = p_{ji}(h)
E[\Phi(h; 0)^{[m]} \mid \sigma_h = i, \sigma_0 = j]$ is Schur stable for every
$h>0$. We {continuously} extend the domain of $\mathcal A$ to the origin by
letting $\mathcal A(0) = I$. {The continuity of the extended $\mathcal A$ at the
origin can be verified by the Dominant Convergence theorem.} Then, to complete
the proof of the corollary, it is sufficient to show that
\begin{equation}\label{eq:sufficietosho}
\mathcal A(h) = \exp(\mathcal B_\Sigma h)
\end{equation}
for every $h\geq 0$.}

{To prove \eqref{eq:sufficietosho}, we need to show that 
\begin{equation}\label{eq:semigroup}
\mathcal A(h)\mathcal A(h') = \mathcal A(h+h') \text{  for all  } h, h'\geq 0
\end{equation}
and
\begin{equation}\label{eq:derivative(0)}
\mathcal A'(0) = \mathcal B_\Sigma. 
\end{equation}
Let us show \eqref{eq:semigroup}. Notice that, by 
\eqref{eq:one-step}, the trajectory $x$ of $\Sigma$ satisfies
\begin{equation}
E[e_{\sigma_{t+h}} \otimes x(t+h)^{[m]}] = \mathcal A(h) E[e_{\sigma_{t}}
\otimes x(t)^{[m]}] \label{eq:one-step:MJLS}
\end{equation}
for all $t\geq 0$ and $h\geq 0$. This in particular implies $\mathcal A{(h+h')}
(e_{\sigma_{0}} \otimes x_0^{[m]}) = \mathcal A{(h)} \mathcal A{(h')}
(e_{\sigma_{0}}  \otimes x_0^{[m]})$ for all $\sigma_0 \in \{1, \dotsc, N\}$ and
$x_0 \in \mathbb{R}^n$. Since the set $\{e_i  \otimes x_0^{[m]} : \text{$i \in
\{1 \dotsc, N\}$ and $x_0\in\mathbb{R}^n$}\}$ spans the whole space
$\mathbb{R}^{Nn_m}$ by \cite[Lemma~1.4]{Ogura2013f}, we obtain
\eqref{eq:semigroup}.} \label{eq:theIdentity}

Then let us show \eqref{eq:derivative(0)}. We compute $\mathcal
A_{ij}'(0)$, the $(i, j)$-block of $\mathcal A'(0)$, for each pair $(i, j)$.
First assume $i\neq j$. Since \changed{$\lim_{h\to 0}E[\Phi(h; 0)^{[m]} \mid
\sigma_h = j, \sigma_0 = i] = I$}{$\lim_{h\to 0}E[\Phi(h; 0)^{[m]} \mid \sigma_h
= i, \sigma_0 = j] = I$,} by the Dominated convergence theorem, we can show
$\mathcal A_{ij}'(0) = \lim_{h\to 0}(p_{ji}(h)I - O)/h = \lambda_{ji} I$. Next
assume $i=j$. Let $\gamma$ denote the number of transitions of $\sigma$ on the
interval $[0, h]$ when $\sigma_0 = i$. Then the event $\{\sigma_h = i, \sigma_0
= i, \gamma=1\}$ has probability zero. Moreover, using the big-O asymptotic
notation, we can show that
\begin{equation*}
\begin{multlined}[.9\linewidth]
E[\Phi(h; 0)^{[m]} \mid \sigma_h = i, \sigma_0 = i, \gamma=\ell] 
\\
=  \begin{cases}
e^{(A_i)_{[m]} h},  & \text{for } \ell = 0, 
\\
I + O(h), & \text{for } \ell \geq 2, 
\end{cases}
\end{multlined}
\end{equation*}
$P(\gamma = 0) = e^{\lambda_{ii}h}$, and $P(\gamma \geq 2) = O(h^2)$.
Therefore
\begin{equation*}
\begin{aligned}
\frac{\mathcal A_{ii}(h) - \mathcal A_{ii}(0)}{h}
&=
\frac{p_{ii}(h)e^{(A_i)_{[m]}h} - I}{h} + O(h)
\\
&\to
(A_i)_{[m]} +\lambda_{ii}I
\end{aligned}
\end{equation*}
as $h\to 0$. The above argument proves \eqref{eq:derivative(0)} and, therefore,
completes the proof of the corollary.
\end{pf}

\subsection{Markov jump linear systems with periodic mode observation}
\label{subsec:periodic}

{In this subsection, we present an example of a system to which none of the results in Corollaries~\ref{cor:regen}\nobreakdash--\ref{cor:Markov} are applicable. Consider the constant-gain state-feedback control of the Markov jump linear system $dx/dt = A_{P, r_t} x(t) + B_{P, r_t} u(t)$, where $A_{P,1}$, $\dotsc$, $A_{P,N} \in \mathbb{R}^{n\times n}$, $B_{P,1}, \dotsc, B_{P,N} \in \mathbb{R}^{n\times p}$, and $r$ is a time-homogeneous Markov process with state space~$\{1, \dotsc, N\}$ and infinitesimal generator~$Q$. If the controller can measure $r$ at any time instant, one can consider the feedback control
\begin{equation}\label{eq:contObsv}
u = K_r x
\end{equation}
with mode-dependent gains $K_i \in \mathbb{R}^{p\times n}$ ($i=1, \dotsc,
N$). It is well known~\cite{Costa2013} that, under this ideal situation, one can
find feedback gains $K_i$ that stabilize the closed-loop system in the
mean-square sense by solving linear matrix inequalities.}

Following the formulation of \citeasnoun{Cetinkaya2014b} in discrete-time, we here
consider a more realistic situation where only the periodic samples
$\{r_{kh}\}_{k\geq 0}$ with a known sampling period $h>0$ are available to the
controller. Precisely speaking, we assume that the feedback control takes the
form
\begin{equation}\label{eq:peri-sfb}
u = K_q x
\end{equation}
where the stochastic process $q$ is defined by $q_t = r_{kh}$ if $kh \leq t<
(k+1)h$ for all $t\geq 0$ and $k\geq 0$. We emphasize that, though we assume
that only periodic samples of $r$ are available, the
infinitesimal generator $Q$ is assumed to be known before the feedback control is designed. \label{alreadyIdentified}Defining $\sigma = (r,q)$ and
$A_\sigma = A_{P,r} + B_{P,r} K_q$, we can write the closed-loop equation in the
form~\eqref{eq:def:Sigma}, which we denote by $\Sigma_h$. Since $r$ is
time-homogeneous and Markovian, we can see that $\sigma$ is a Markov
regenerative process with an associated embedded Markov renewal process $\{(r_{kh},
kh)\}_{k\geq 0}$. The function $\pi$ in \ref{item:determined }
maps the pair $(i, j) \in \{1, \dotsc, N\}^2$ to $i$. Then,
using Theorem~\ref{theorem:main}, we can prove the following characterization of
the exponential mean stability of $\Sigma_h$:

\begin{cor} \label{cor:2}
{Assume that $m$ is even. For each $j \in \{1, \dotsc, N\}$, let $\mathcal
B_{j} = Q^\top \otimes I_{n_m} + \bigoplus_{i=1}^N (A_{(i,j)})_{[m]}$. Define
\begin{equation}\label{eq:calAh}
\mathcal A_h  
= 
\sum_{j=1}^N  \exp(\mathcal B_j h) ((e_je_j^\top) \otimes I_{n_m}).
\end{equation}
Then, $\Sigma_h$ is exponentially $m$th mean stable if and only if $\mathcal
A_h$ is Schur stable.}
\end{cor}

\begin{pf}
{Since $m$ is even, condition \ref{item:evenORnon-neg} is satisfied.
Moreover, because $\tau_{k+1}-\tau_k = h$, condition \ref{item:hk<T} as well as
\eqref{eq:tauk+1-tauk>0} hold true. Therefore, by Theorem~\ref{theorem:main}, it
is sufficient to show that the matrix $\mathcal A$ given by \eqref{eq:calAij}
equals $\mathcal A_h$. To compute $\mathcal A_{ij}$, observe that $\sigma$
confined in $[0, h)$ with the condition {$r_0 = j$} is a time-homogeneous Markov
process with state space $\{(1, j), \dotsc, (N, j)\}$ and infinitesimal
generator $Q$. Therefore, from \eqref{eq:sufficietosho}, $\mathcal A_{ij}$
equals the $(i, j)$-block of the matrix $\exp(\mathcal B_{j} h)$. Hence, the
$j$-th block-column of $\mathcal A$ equals that of $\exp(\mathcal B_{j} h)$.
This proves $\mathcal A = \mathcal A_h$, as desired.}
\end{pf}

\begin{rem}\label{rem:cannotByStudied}
{Note that we can use neither Corollaries~\ref{cor:regen} nor \ref{cor:semi-} to
analyze the stability of $\Sigma_h$. First, it is shown in
\citeasnoun[Example~7]{Ogura2014d} that $\sigma$ cannot be a semi-Markov process.
Moreover, though $\sigma$ can be realized as a regenerative process, the
realization in~\citeasnoun[Example~7]{Ogura2014d} does not satisfy \ref{item:hk<T} by
the following reason. In the embedded renewal process $(\theta, \tau)$ of the
realization, the renewal times $\{\tau_k\}_{k\geq 0}$ are given as the sampling
times $t = kh$ ($k=0, 1, \dotsc$) such that $r(t) = r_0$. However, the
difference $\tau_{k+1}- \tau_k$ can take an arbitrary large number and, hence,
cannot be bounded by a uniform and finite number.}
\end{rem}

Let us see an example. Consider the Markov jump linear system with the following
coefficient matrices:
\begin{equation*}
\begin{gathered}
A_1 = \begin{bmatrix}
0&0&0\\
0&-0.545 & 0.626
\\
0&-1.570 & 1.465
\end{bmatrix}, \ 
A_2 = \begin{bmatrix}
0 & 0 & 0
\\
0 & -0.106 & 0.087
\\
0 & -3.810 & 3.861
\end{bmatrix},\\
A_3 = \begin{bmatrix}
1,80 & -0.3925 & 4.52 
\\
3.14 & 0.100 & -0.28
\\
-19.06 & -0.148 & 1.56
\end{bmatrix}, 
\\
B_1 = \begin{bmatrix}
0 \\-0.283 \\ 0.333
\end{bmatrix},\ 
B_2 = \begin{bmatrix}
0 \\ 0 \\ 0.087
\end{bmatrix},\ 
B_3 = \begin{bmatrix}
-0.064 \\ 0.195 \\ -0.080
\end{bmatrix}, 
\end{gathered}
\end{equation*}
and the infinitesimal generator
\begin{equation*}
Q = \begin{bmatrix}
-0.53 & 0.32 & 0.21\\
0.50 & -0.88 & 0.38\\
0.40 & 0.13 & -0.53
\end{bmatrix}.
\end{equation*}
This system, taken from \citeasnoun{Blair1975}, models a certain economic system. We
denote by $\Sigma_0$ the closed-loop system when the classical feedback control
\eqref{eq:contObsv} is applied to this system. In \citeasnoun{Costa2013}, the feedback
gains that stabilize $\Sigma_0$ (in the mean square sense)  with the minimum
$H^2$ norm are obtained as
\begin{equation*}
\begin{aligned}
K_1 &= \begin{bmatrix}
2.0343 & 14.5181 & -23.5917
\end{bmatrix}, 
\\
K_2 &= \begin{bmatrix}
1.0187 & 73.0961 & -78.7596
\end{bmatrix}
,
\\
K_3 &= \begin{bmatrix}
93.6651 & -11.4921 & 11.6875
\end{bmatrix}.
\end{aligned}
\end{equation*}

\begin{figure}[tb]
\centering
\includegraphics[width=0.9\linewidth]{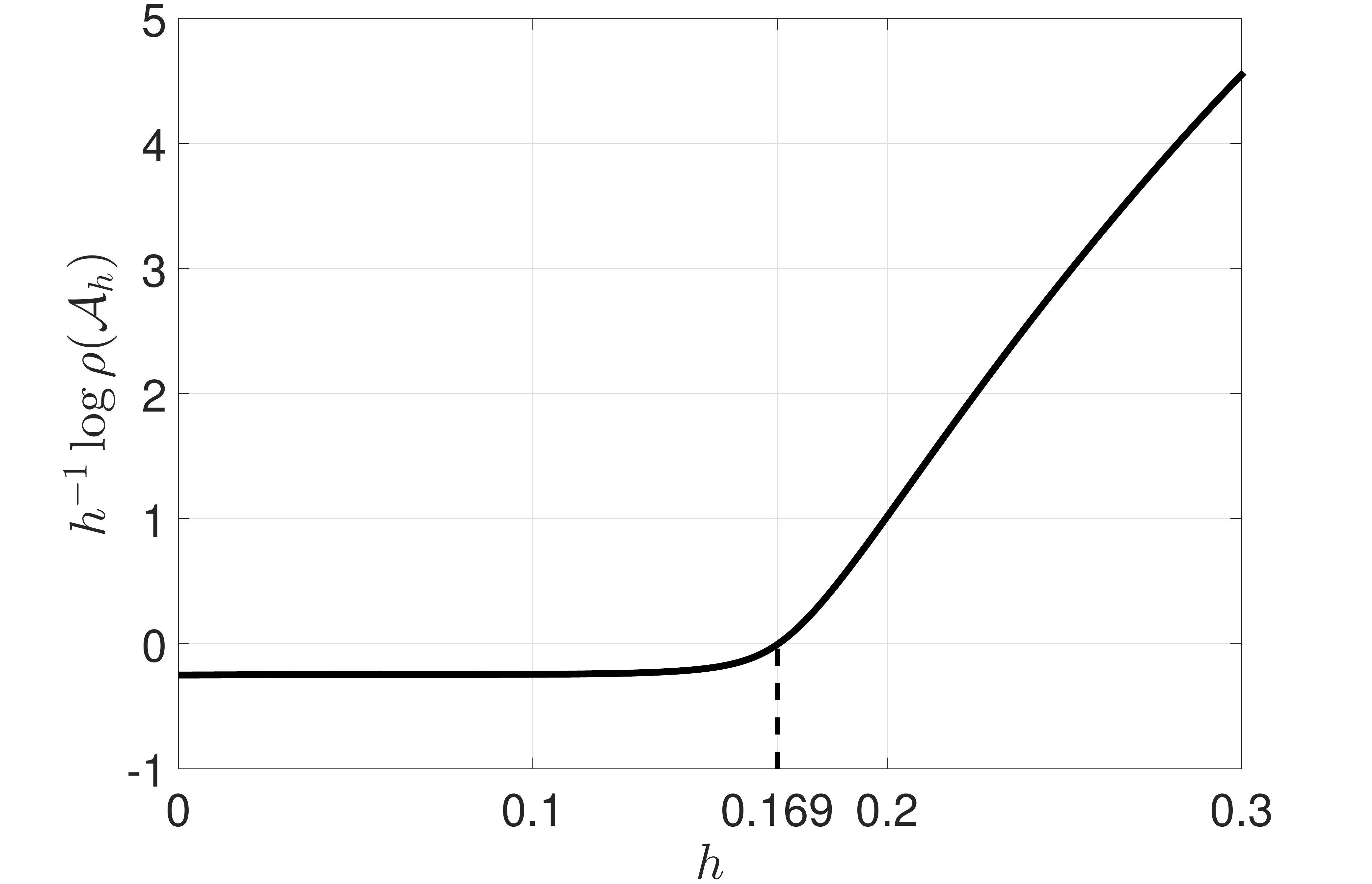}
\caption{$h^{-1}\log(\rho(\mathcal A_h))$ versus $h$}
\label{fig:untitled}
\end{figure}

{We use Corollary~\ref{cor:2} to investigate how the stability property of $\Sigma_{0}$ is altered when the feedback control~\eqref{eq:contObsv} is replaced with the one in \eqref{eq:peri-sfb} based on periodic observations. For $m=2$ and period $h = (0.001)\ell$ ($\ell=1, \dotsc, 300$), we compute the spectral radius of $\mathcal A_h$ given in Corollary~\ref{cor:2}. In Fig.~\ref{fig:untitled}, we show the graph of $h^{-1}\log(\rho(\mathcal A_h))$ as $h$ varies. \label{computationDetail} From the corollary and the graph, we determine that the system $\Sigma_h$ is mean square stable if and only if $0<h<0.169$ [years].} This implies that, to guarantee the stability of the controlled economic system, $r$ must be sampled with a period less than about 2 months. {It is interesting to observe that, as $h\to 0$, the quantity $h^{-1}\log(\rho(\mathcal A_h))$ becomes close to $-0.250$, the maximum real part of the eigenvalue of the matrix $\mathcal B_{\Sigma_{0}}$ that characterizes the stability of the original system $\Sigma_{0}$ by Corollary~\ref{cor:Markov}. This shows that, in the limit of $h\to 0$, the stability of the original closed-loop system $\Sigma_{0}$ is ``recovered'' by $\Sigma_h$.}

\begin{figure}[tb]
\begin{minipage}[b]{\linewidth}
\centering
\includegraphics[width=0.9\linewidth]{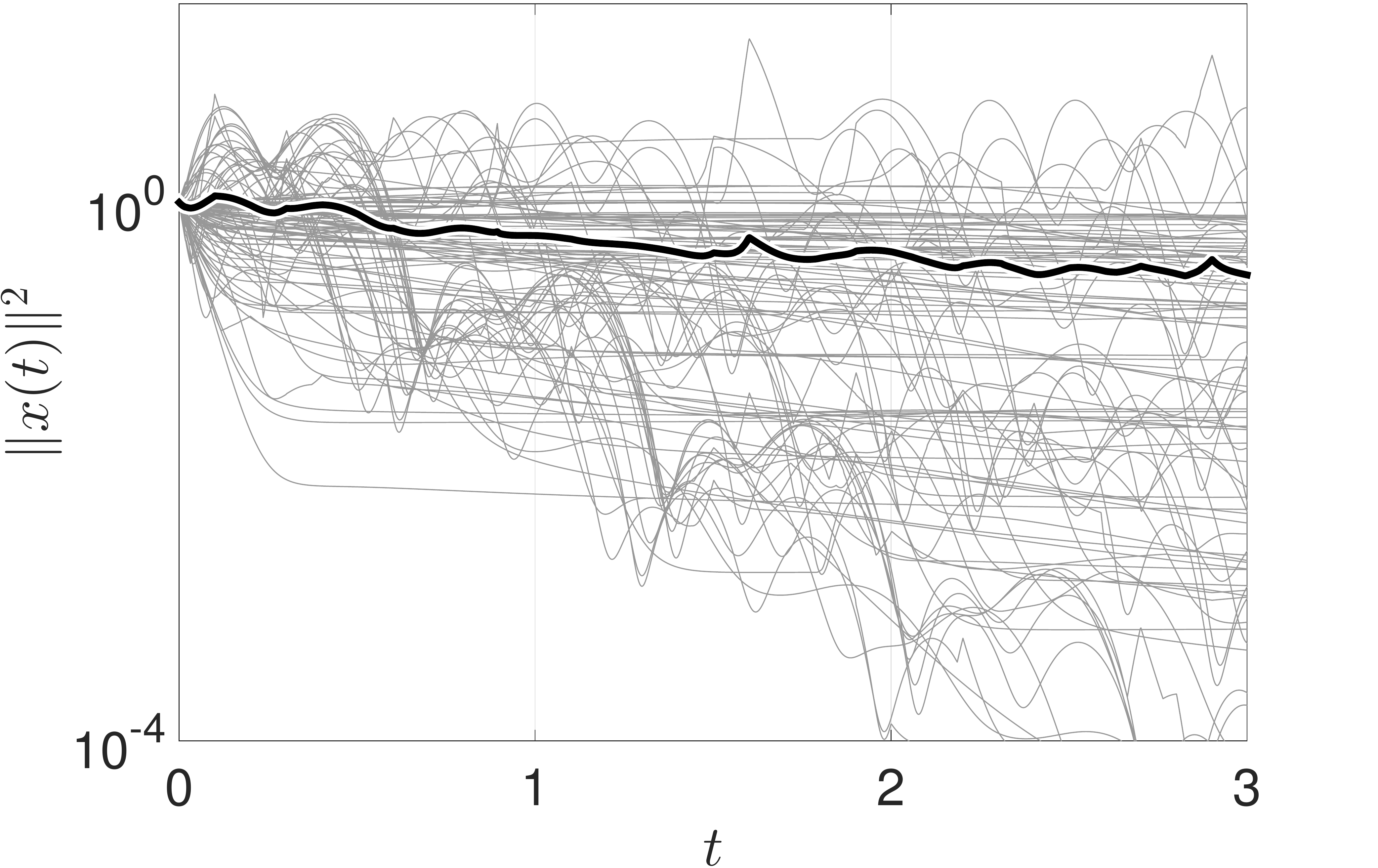}
\subcaption{{$h = 0.1$ (stable)}}
\label{fig:stable}
\end{minipage}%
\vspace{.5cm}
\\
\begin{minipage}[b]{\linewidth}
\centering
\includegraphics[width=0.9\linewidth]{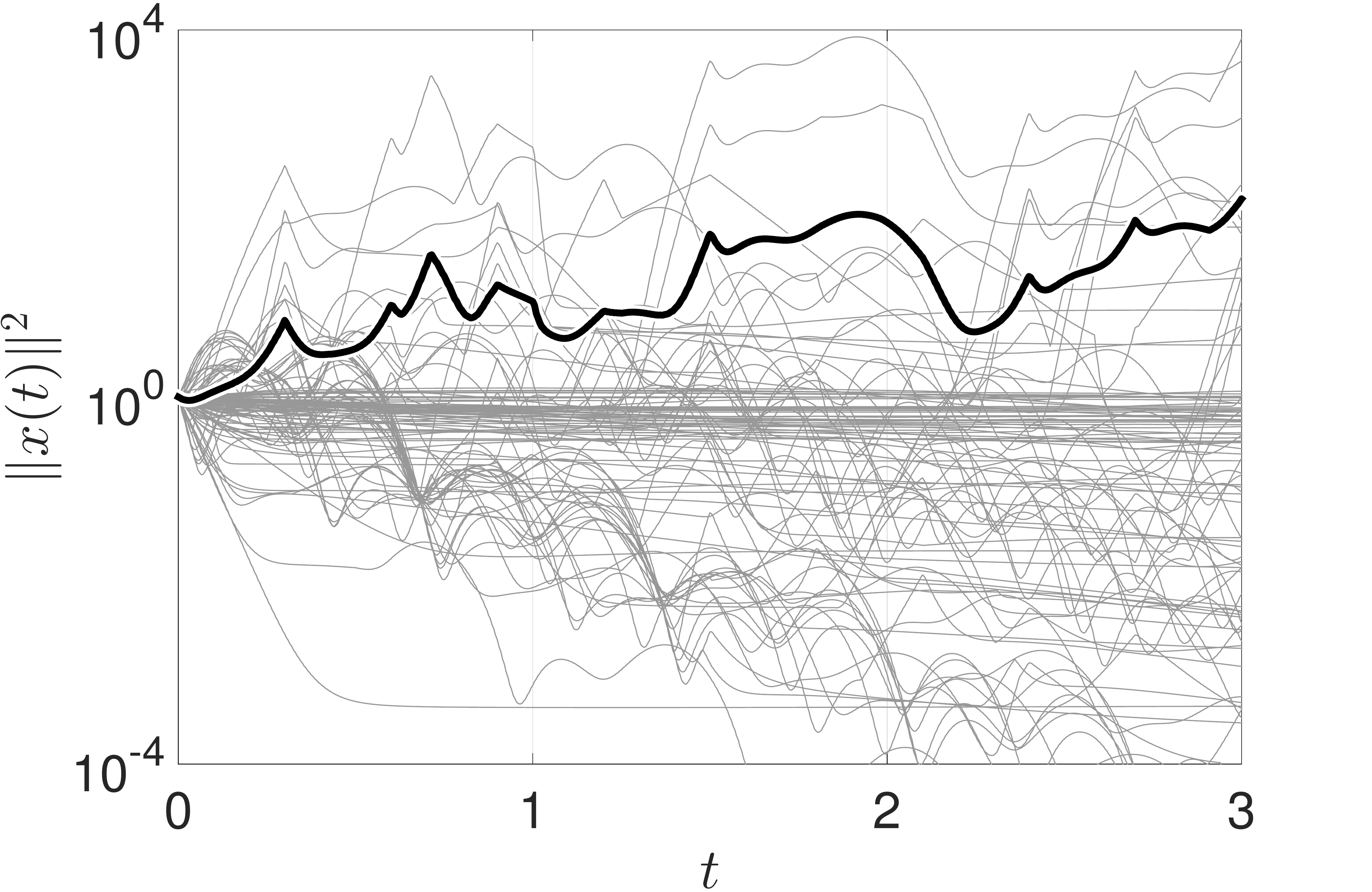}
\subcaption{{$h = 0.3$ (unstable)}}
\label{fig:unstable}
\end{minipage}
\caption{{Sample paths of $\norm{x(t)}^2$ (thin lines) and their averages
(thick lines) for $h=0.1$ and $0.3$}}\label{fig:sample}
\end{figure}

\begin{figure}[!t]
\begin{minipage}[b]{\linewidth}
\centering
\includegraphics[width=.9\linewidth]{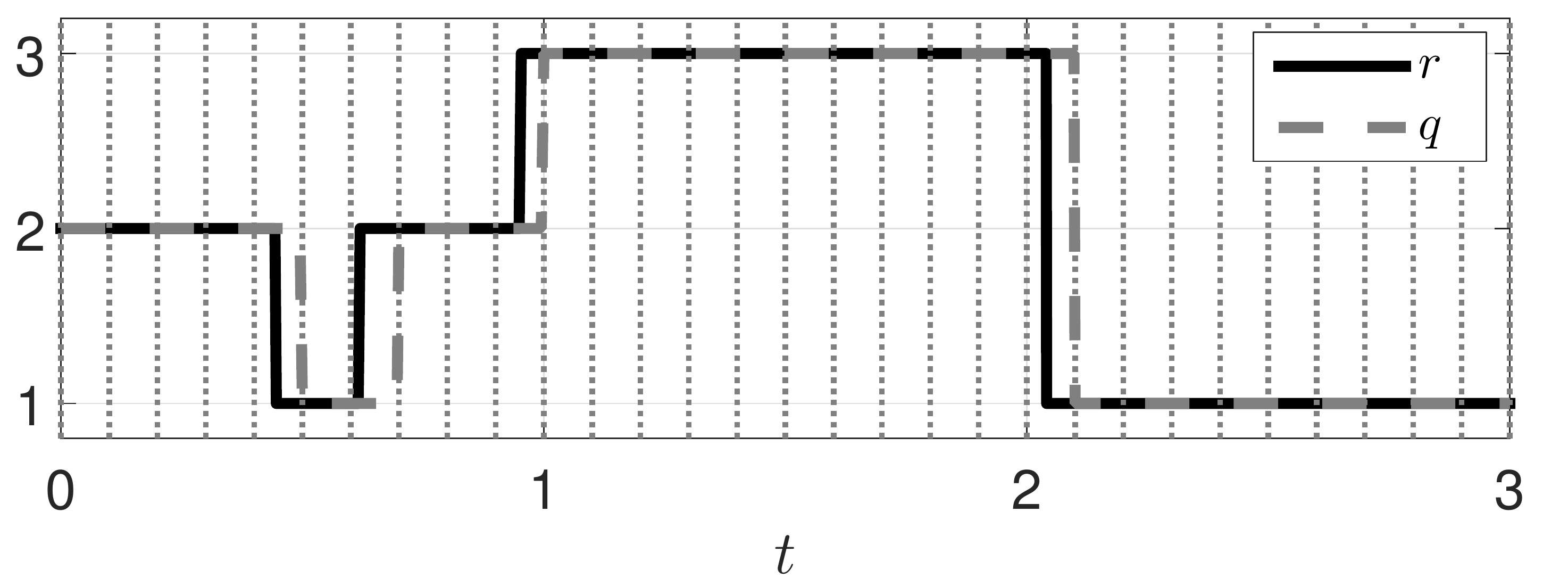}
\subcaption{{$h = 0.1$ (stable)}}
\vspace{.3cm}
\end{minipage}
\\
\begin{minipage}[b]{\linewidth}
\centering
\includegraphics[width=.9\linewidth]{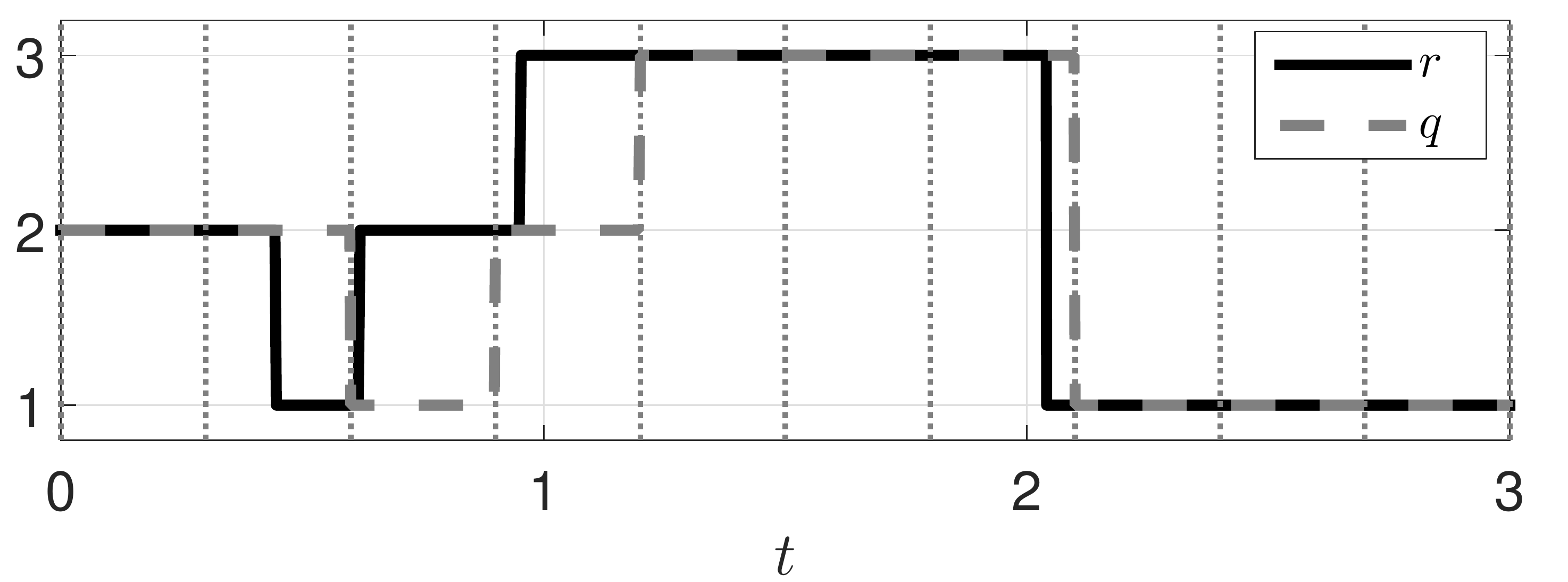}
\subcaption{{$h = 0.3$ (unstable)}}
\end{minipage}
\caption{{Sample paths of $\sigma = (r,q)$ for $h=0.1$ and $0.3$. Solid lines, dashed lines, and dotted lines show $r$, $q$, and sampling instants, respectively.}}\label{fig:samplePathofSigma}
\end{figure}

In Figs.~\ref{fig:stable} and \ref{fig:unstable}, we show $100$ sample paths of $\norm{x(t)}^2$ and their sample averages when $h = 0.1$ and $h = 0.3$, respectively. For the generation of each sample path, we take $x_0$ randomly from the unit sphere in $\mathbb{R}^3$ and $r_0$ from the uniform distribution on $\{1, 2, 3\}$. We can see that mean square stability is achieved with sampling period $h=0.1$, while the closed-loop system exhibits instability for $h=0.3$. {Finally, for each value of $h$, we show sample paths of $\sigma = (r,q)$ in Fig.~\ref{fig:samplePathofSigma}. We notice that the change of the values of $q$ occurs only at sampling instants, shown by the dotted lines.}

\section{Conclusion}

In this paper, we have investigated the mean stability of Markov regenerative
switched linear systems. The class of switched linear systems contain a wide
variety of important stochastic switched linear systems that have appeared in
the literature. We have shown that the mean stability of a Markov regenerative
switched linear system is characterized by the spectral radius of a matrix
arising from its transition matrix. A numerical example was presented to
illustrate the obtained result.

\end{document}